\documentclass[twocolumn,prb,showpacs,a4paper]{revtex4}
\usepackage{dcolumn}
\usepackage{bm}
\usepackage[T1]{fontenc}
\usepackage{graphicx}
\usepackage{multirow}
\usepackage{color}

\begin{document}

\title{First-principles study of the adsorption of MgO molecules on a clean Fe(001) surface}
\author{Damian Wi\'snios$^{1,2}$}
\author{Adam Kiejna$^{2}$}\email{kiejna@ifd.uni.wroc.pl}
\author{J\'ozef Korecki$^{1,3}$}
\affiliation{$^{1}$Faculty of Physics and Applied Computer Science, AGH University of Science and Technology, al. Mickiewicza 30, 30-059 Krak\'ow, Poland}
\affiliation{$^{2}$Institute of Experimental Physics, University of Wroc{\l}aw, pl. M. Borna 9, 50-204 Wroc{\l}aw, Poland}
\affiliation{$^{3}$Jerzy Haber Institute of Catalysis and Surface Chemistry, Polish Academy of Sciences, ul. Niezapominajek 8, 30-239 Krak\'ow, Poland}
\date{\today}

\begin{abstract} 
The adsorption of MgO molecules on a Fe(001) surface was studied using density functional theory and projector augmented wave methods. The energetically most favored configurations for different adsorption sites considered were identified. The most preferable adsorption geometry is when the MgO molecules are parallel to the surface, with Mg in the interstitial site and O in on-top of the Fe atom. During the adsorption of subsequent MgO molecules in this geometry, a sharp, non-oxidized interface is formed between the MgO adlayer and Fe(001) surface. The adsorption of MgO perpendicular to the surface, with oxygen incorporated in the topmost Fe layer is less probable, but it may lead to the formation of the FeO layer when stabilized with an excess of oxygen atoms. Structural, electronic and magnetic properties of both interface types were examined for the MgO coverage from 1/9 to 1 monolayer (ML). Electronic and magnetic properties are sensitive to the MgO coverage. For lower coverage of MgO, clear hybridization between the Fe $3d$ and O $2p$ states is shown. The average magnetic moment of the surface Fe atoms is reduced with coverage, achieving 2.78 $\mu_{\rm B}$ for 1 ML of MgO.
\end{abstract}


\maketitle

\section{Introduction}

Thin-film multilayer systems have attracted considerable research interest for many years due to their technological application in magnetic tunnel junctions, magnetic memories (MRAM) or spin valves used in high-density magnetic recording devices. 
In particular, thin-film systems containing Fe and MgO layers have been investigated by many authors due to their fundamental properties, such as enhanced magnetic moments,~\cite{Koyano1988} interlayer exchange coupling,~\cite{Katayama2006, Faure2002} huge tunneling  magnetoresistance~\cite{Mathon2001,Butler2001} (TMR), and perpendicular magnetic anisotropy.~\cite{Yang2011,Lambert2013,Balogh2013,Koziol2013}

A small lattice mismatch between MgO(001) and Fe(001) surface unit cells, $(d_{\rm MgO}-\sqrt{2}d_{\rm Fe})/d_{\rm MgO}\approx 4\%$, makes favorable conditions for the epitaxial growth of both Fe on MgO (Ref.~\onlinecite{Lawler1997,Subagyo1999}) and MgO on Fe (Ref.~\onlinecite{Vassent1996,Popova2002}) with well defined orientation relations: Fe(001)|[100] $\parallel$ MgO(001)|[110]. 
However, the interfacial electronic and chemical structure of the Fe/MgO interface becomes more complex when considering such effects as Fe-O bond formation and oxidation.~\cite{Palomares2005,Bonell2009} This could strongly reduce the Fe spin polarization at the interface and thus it makes a particular interface structure essential for the TMR effect.~\cite{Bose2008}
Experimentally, many different, apparently contradicting structures at the Fe/MgO and MgO/Fe interfaces were reported, ranging from nearly iron-oxide-free~\cite{Fullerton1995,Luches2005,Plucinski2007,Colonna2009} to the formation of an FeO layer.~\cite{Meyerheim2002,Oh2003,Tusche2006} 
The observed differences originate from preparation conditions, either oxygen-deficient (deposition from bulk MgO) or oxygen-rich (reactive deposition of Mg in oxygen atmosphere), as well as the deposition sequence (Fe on the MgO surface or MgO monolayers on Fe). In particular, the interfaces between the MgO and Fe layers (Fe on MgO and MgO on Fe) were experimentally examined by conversion electron M\"{o}ssbauer spectroscopy (CEMS).~\cite{Mlynczak2013}
Under oxygen-deficient conditions, the coexistence of both oxidized and nonoxidized interfaces was  confirmed, and interface properties were found to be highly sensitive to the concentration of defects.~\cite{Mlynczak2013} Different methods have been applied to prevent the formation of FeO layers under oxygen-rich conditions, including modified reactive deposition,~\cite{Tekiel2013} deposition of metallic Mg atoms on the Fe(001) substrate and subsequent annealing of the sample under O$_2$ exposure,~\cite{Dugerjav2011} or a metallic Mg buffer-layer formation followed by reactive Mg deposition.~\cite{Parkin2007} 
On the theory side, a full-potential linearized augmented-plane-wave study of monolayers of Fe on MgO(001) has shown~\cite{Freeman1991} that there is almost no interaction between Fe and MgO. More recent plane-wave DFT calculations \cite{Yu2006,Jeon2011} of MgO monolayers on a Fe(001) substrate confirmed that the MgO films of one to three monolayers (MLs) interact only weakly with the Fe substrate. At the oxidized MgO/FeO/Fe(001) interfaces, an increase of the magnetic moments of Fe atoms near the interface and a reduction of the work function with respect to the clean Fe(001) surface were reported.~\cite{Yu2006,Jeon2011}
	
The first stage of growth, i.e.\ adsorption, plays a significant role in the structure of multilayer systems. Understanding of this initial process allows for an understanding of the nucleation and growth of three-dimensional structures. The MgO substrate is commonly used due to its simple atomic structure and its relevance to many experimental works on model systems.~\cite{Boubeta2003} Experimental~\cite{Urano1988} and theoretical studies~\cite{Yu2005,Kim2006} of Fe adsorption on defect-free MgO(001) surface showed the strongest adsorption binding for Fe adatoms on-top of the surface oxygen atoms. Surprisingly, there are no theoretical works considering the adsorption of single MgO molecules on metal surfaces. The equilibrium structures of the MgO layers on Fe(001) were optimized with an \textit{a priori} assumption that the oxygen preferably adsorbs at the on-top sites. However, the calculation by Beltr\'an \textit{et al.}~\cite{Beltran2012} suggested that under low MgO coverage, Fe binds to Mg rather than to O.

Simulations of MgO adsorption on Fe(001) should account for different deposition methods. The reactive deposition of metallic Mg in a molecular oxygen atmosphere should be avoided because it leads to direct oxidation of the Fe surface layer.~\cite{Oh2003} For the preferred deposition of MgO from the bulk, one should consider the adsorption of MgO molecules and/or of the products of their dissociations, i.e. co-adsorption of atomic (ionic) oxygen and magnesium. 
In the former case, one can refer to the adsorption of atomic oxygen on iron surfaces,~\cite{Blonski2005} which confirmed the experimental observation~\cite{Leygraf1973} that the interstitial (fourfold hollow) position is the most stable oxygen adsorption site on the Fe(001) surface. 
Vassent \textit{et al.}~\cite{Vassent2000} experimentally demonstrated that when MgO is deposited from a bulk target by electron beam bombardment, the resulting beam is composed mainly of atomic Mg and O, which recombine on the Fe(001) substrate. However, adsorption of the MgO molecules should also be considered, especially when an alternative deposition method is used (e.g. pulsed laser deposition or thermal evaporation). A simple mass-spectroscopy experiment performed for the purpose of the present study demonstrated that when MgO is evaporated from a thermal source, the vapor flux, contains, aside from atomic O and Mg, at least 15\% of the MgO molecules.~\cite{Giela2014} The real contribution of the MgO molecules can be considerably higher when considering the dissociation of the MgO molecules in the quadrupole mass spectrometer.
	
In this paper, we report  DFT calculations of the adsorption of MgO molecules on the Fe(001) surface. Starting from a single MgO molecule in a large surface unit cell and by considering different adsorption sites and molecule orientations, the most energetically favorable adsorption geometries were determined. Subsequently, by increasing the coverage to the complete MgO monolayer, we were able to simulate low-temperature deposition of MgO onto the clean Fe(001) surface under ultrahigh-vacuum conditions. We have examined the energetics and electronic properties of the ensuing structures, which are related to the first stages of MgO growth on Fe(001). 

\section{Calculation method}	

The calculations presented in this work were carried out using the Vienna \textit{ab initio} simulation package (VASP) based on density functional theory (DFT).~\cite{Kresse1993,Kresse1996} The exchange-correlation energy was described within the Perdew-Burke-Ernzerhof (PBE) version~\cite{Perdew1996} of the spin-polarized generalized gradient approximation (GGA).~\cite{Perdew1992,Vosko1980} The electron-ion interactions were represented by the projector augmented-wave (PAW) potentials~\cite{Blochl1994} with the $3d^74s^1$, $3s^2$, and $2s^2 2p^4$ states considered as the valence states of the Fe, Mg, and O atoms, respectively. For the plane-wave-basis set, a cutoff energy of 500~eV was applied based on the series of tests. 
The $k$-space integrations for the bulk bcc Fe were performed using a $12 \times 12 \times 12$ special $k$-point mesh   generated by the Monkhorst-Pack method.~\cite{Monkhorst1976} The partial occupancies were treated by using the first-order Methfessel-Paxton method~\cite{Methfessel1989} with a Fermi surface smearing of 0.2~eV.
This computational setup yielded calculated properties of bulk bcc iron that were in good agreement with experiment and previous theoretical works. The optimized lattice constant of 2.832~\AA\ for ferromagnetic bcc Fe differs from the experimental value (2.866~\AA)\cite{Herper1999} by only approximately 1\% and agrees well with the values reported in other theoretical works.~\cite{Jeon2011,Stixrude1994,Kresse1999,Blonski2007,Shimada2010,Zhong1993}
The bulk modulus, obtained by fitting the Murnaghan equation of state to the energy dependence on the lattice parameter, is 179 GPa, which is slightly larger than the measured value of 172 GPa,~\cite{Jephcoat1986} and it is close to the other theoretical GGA~\cite{Stixrude1994,Kresse1999,Blonski2007} and the local density approximation (LDA)~\cite{Zhong1993} results. 
The calculated cohesive energy of 4.943 eV is 15\% larger than the experimental value (4.28~eV),~\cite{Kittel1986} but it is similar to previous results obtained by other authors using both pseudopotential~\cite{Zhong1993} and full-potential~\cite{Sato2009,Tang2002} first-principles methods. 
The computed magnetic moment of 2.19 $\mu_{B}$ is also in good agreement with the experimental bulk value of 2.22 $\mu_{B}$~\cite{Kittel1986} and other calculations.~\cite{Stixrude1994,Kresse1999, Shimada2010, Blonski2007, Hugosson2013,Zhong1993}

The Fe(001) surface was modeled by a slab of 9 Fe layers separated from its periodic replicas by a thick vacuum region of 22~\AA. An adequate number of Monkhorst-Pack $k$-points has been selected for each system depending on the size of the unit cell:  $12 \times 12 \times 1$ for the clean Fe(001) surface $1 \times 1$ cell and $4 \times 4 \times 1$ for MgO adsorption in the $3 \times 3$ surface unit cell. MgO molecules were adsorbed on both sides of the slab. 
The positions of all atoms were optimized until the forces exerted on each atom were less than 0.01~eV/\AA.

The adsorption energy was calculated from the total energy difference:
\begin{equation}
 E_{\rm ad}=-\frac{1}{N}(E_{\rm MgO/Fe(001)}-E_{\rm Fe(001)}-NE_{\rm MgO}),
\end{equation}
where $E_{\rm MgO/Fe(001)}$ and $E_{\rm Fe(001)}$ represent the energy of the slab with adsorbed MgO molecules and the total energy of the clean slab, respectively. $N$ is the number of molecules adsorbed on both sides of the slab, and $E_{\rm MgO}$ is the energy of the free MgO molecule calculated in a large rectangular box with dimensions of 11 $\times$ 12 $\times$ 13 \AA. The calculated bond length of a free MgO molecule is 1.75 \AA. 

\section{Results and discussion}
\subsection{Clean Fe(001) surface}

The calculated basic properties of the clean Fe(001) surface, lattice relaxation, surface energy, work function, and magnetic moments on the surface atoms create a solid reference system for MgO adsorption. 
The obtained values of relaxation, $\Delta_{ij} = (d_{ij}-d)/d$, of surface atomic layers distance, $d_{ij}$, between subsequent layers $i$ and $j$, where $d$ is the interplanar distance in bulk Fe, fall well (Table \ref{tab:relax}) within the range of values determined by calculations applying the same~\cite{Jeon2011,Hugosson2013} or a similar flavor of GGA-PW91.~\cite{Blonski2007,Spencer2002} 
The computed surface energy, $E_{\rm surf}=2.52$~J/m$^2$, is between two experimental results, 2.41 J/m$^2$ (Ref.~\onlinecite{Tyson1977}) and 2.55 J/m$^2$ (Ref.~\onlinecite{Miedema1978}), and is of a similar magnitude as the result reported by other authors.~\cite{Jeon2011,Blonski2007,Shimada2010} This energy was determined from the expression $E_{\rm surf}=\frac{1}{2A}(E_{\rm n}-nE_{\rm B})$, where $E_{n}$ is the total energy of the slab, $n$ is the number of layers in the slab, $A$ is the surface area, and  the $1/2$ factor accounts for the two surfaces of the slab. $E_{\rm B}$ is the energy of the bulk layer calculated as the difference between the total energy of the $(n+1)$- and $n$-layer slabs. 
The work function of 3.86 eV, calculated as the difference between the electrostatic potential energy in the vacuum region and the Fermi energy of the slab, is in perfect agreement with earlier GGA calculations.~\cite{Jeon2011,Blonski2007,Hugosson2013} However, it is substantially less than the experimental values of 4.67 eV (Ref.~\onlinecite{Michaelson1977}) and 4.24 eV (Ref.~\onlinecite{Paggel2002}).
The magnetic moments on surface and subsurface Fe atoms of the Fe(001) slab, due to their lower coordination, are enhanced to 2.94 and 2.33 $\mu_{\rm B}$, respectively. They agree well with previous  \cite{Jeon2011, Shimada2010, Hugosson2013, Blonski2007} GGA results. For the deeper layer atoms, the magnetic moments approach the value for bulk iron of 2.19 $\mu_{\rm B}$.  

\subsection{MgO molecules on the clean Fe(001) surface}

\subsubsection{Geometry and energetics}

\begin{figure}[]
\centering
\includegraphics[width=3cm]{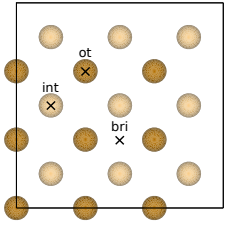}
\caption{(color online) Adsorption sites (crosses) for MgO molecules in the $3\times3$ surface unit cell representing the Fe(001) surface: ot - on-top, int - interstitial, bri - bridge site. The brown and light brown balls correspond to the first and second Fe layers, respectively. }
\label{fig:site}
\end{figure}

\begin{table}[]
\caption{Adsorption energy $E_{\rm ad}$ for a single MgO molecule adsorbed on the Fe(001) surface for all initially considered geometries. For the parallel geometry, the first part of each term describes the location of the Mg and the second part describes the location of the O atom. For the perpendicular geometry, the first part of the term describes the atom that bonds with the surface and the second part determines the location of the molecule.}
\label{tab:eads}
\begin{ruledtabular}
\begin{tabular}{cccc}
\multicolumn{2}{c}{Parallel geometry} & \multicolumn{2}{c}{Perpendicular geometry} \\ \hline
 Configuration & $E_{\rm ad}$ [eV] & Configuration & $E_{\rm ad}$ [eV]  \\ \hline
 int-ot		&	4.60		&	O-ot		&	3.31	  \\
 ot-int		&	4.37		&	O-int	&	4.19  \\
 bri-ot		&	4.17		&	O-bri	&	3.83  \\
 ot-bri		&	4.14		&	Mg-ot	&	1.38  \\
 bri-int		&	4.40		&	Mg-int	&	0.64  \\
 int-bri		&	4.90		&	Mg-bri	&	1.33	  \\
\end{tabular}
\end{ruledtabular}
\end{table}

MgO molecules were adsorbed on the relaxed Fe(001)-oriented slab. Figure~\ref{fig:site} shows the adsorption sites considered for Mg and O atoms of the MgO molecule at the Fe(001) surface: on-top (ot) of the Fe atom, the interstitial (int) four-fold hollow between the four surface Fe atoms and the bridge (bri) site between the two nearest surface Fe atoms. Adsorption geometries tested in the calculations included different possible alignments of the MgO molecules with respect to the surface: parallel, with Mg and O at different positions, and perpendicular, with the Mg or O atom bonding with the substrate. 
This resulted in six orientations of MgO molecules parallel to the surface, termed int-ot, ot-int, bri-ot, ot-bri, bri-int, and int-bri, where the first part of each term in this notation describes the location of the Mg and the second part describes the location of the O atom. Similarly, six configurations of the MgO molecule oriented perpendicular to the surface were considered: three of them bonding with O (O-ot, O-int, O-bri) and the remaining three bonding with the Mg atom to the Fe(001) surface (Mg-ot, Mg-int, Mg-bri). 

\begin{figure}[]
\includegraphics[width=7.0cm]{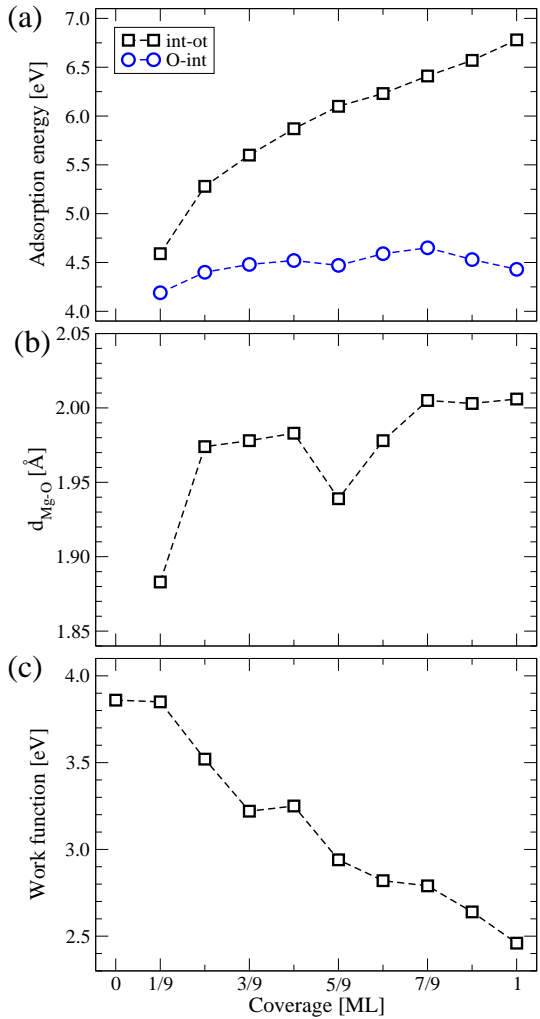}
\caption{(color online) Coverage dependence of adsorption energy (a), Mg-O distance (b) and work function (c) for the most favorable overall int-ot orientation of MgO molecules adsorbed on the Fe(001) surface. For comparison, the adsorption energy plot in (a) also presents the results for the most favorable perpendicular O-int orientation of the molecules.}
\label{fig:ads}
\end{figure}

\begin{figure}[!]
\centering
\includegraphics*[width=6.5cm]{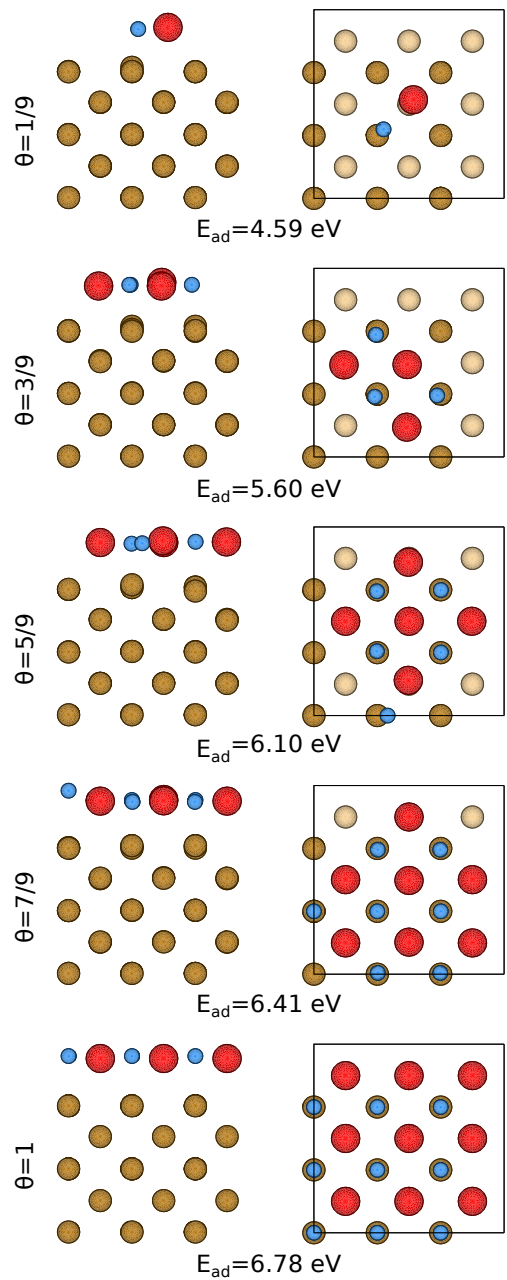}
\caption{(color online) Side and top (right column) views of the most stable final int-ot configuration of MgO molecules adsorbed on the Fe(001) surface, with the Mg atom (red) in the interstitial (int) and the O atom (blue) on-top (ot) of the Fe atom. Medium (brown) balls represent Fe atoms. In the top view, Fe atoms of the second layer are a lighter shade.}
\label{fig:fconf}
\end{figure}

The adsorption energy for a single MgO molecule adsorbed on the $3\times 3$ surface unit cell (a coverage of 1/9 ML) for different adsorption geometries is shown in Table~\ref{tab:eads}. Configurations aligned parallel to the surface have the highest adsorption energy, especially the int-ot and the int-bri systems with values of 4.60 eV and 4.90 eV, respectively.
Generally, the bonding of the single MgO molecule adsorbed perpendicularly is substantially weaker, especially via the Mg ($E_{\rm ad}\approx$1 eV compared to less than 4 eV for the binding via the O atom). The only exception is the O-int configuration, whose adsorption energy of 4.19 eV is comparable with the parallel MgO configuration. In this case, the oxygen atom is at the closest distance to the surface, 0.69~\AA, and its local configuration is similar to that in the FeO(001) monolayer. 

A coverage dependence of the stability of different adsorbate structures is discussed below. Starting from a single MgO molecule on the $3\times3$ surface unit cell, by the subsequent addition of an extra MgO molecule, in all possible arrangements with respect to the previous coverage and preserving the appropriate adsorption configuration, the coverage dependence was followed up to a complete MgO monolayer. In this manner, the lowest-energy arrangement for each degree of coverage was determined.
Only the int-ot configuration with MgO aligned parallel to the surface and the perpendicular O-int configuration led to the experimentally observed structure, i.e. to the pseudomorphic MgO(001)/Fe(001) interface. Other parallel configurations of a single MgO molecule, including int-bri, which shows the strongest adsorption binding for $\mathrm{\Theta}$=1/9 ML, are at higher coverages considerably more weakly bound than the int-ot phase, and they do not form any reasonable (known) structure.
In turn, at higher coverages, the binding of MgO in the perpendicular O-int configuration is distinctly weaker than in the int-ot.
Therefore, and taking into account the highest adsorption energy for higher degrees of coverage, we consider the int-ot system as the most stable configuration.
Therefore, in the following, we concentrate on the parallel int-ot system and some aspects of the perpendicular O-int configuration in which the surface Fe layer is prone to oxidation. 

The adsorption energy as a function of MgO coverage is plotted in Fig.~\ref{fig:ads}(a). In the int-ot configuration, with the Mg atoms in the interstitial sites and the O atoms in the on-top sites, the adsorption energy increases monotonically with the coverage, from 4.60 eV for 1/9 ML to 6.78 eV for 1 ML of MgO. 
For comparison, in Fig.~\ref{fig:ads}(a) we have also plotted the variation of the adsorption energy versus  coverage for MgO molecules adsorbed in the perpendicular O-int configuration, with the O atom closer to the surface. In this case, the MgO binding is weaker than for parallel-oriented MgO molecules. It increases only slightly with increasing coverage, and the difference in the adsorption bond strength for 1 ML of coverage is larger than 2 eV. This shows that when stoichiometric MgO is adsorbed, the formation of the interfacial FeO layer is less favorable. 

The optimized configurations for several coverages of the int-ot-oriented MgO molecule are shown in  Fig.~\ref{fig:fconf}. As can be seen, MgO molecules adsorbed on the surface have a tendency to form compact, symmetric two-dimensional MgO surface clusters. This is in agreement with the experimentally observed tendency of island growth during MgO deposition.\cite{Klaua2001} Any asymmetry in the cluster shape would lead to atom displacements from an ideal adsorption position, as observed for $\Theta$=5/9, where one of the O atoms is clearly shifted with respect to the other atoms due to a smaller number of the nearest neighbours.
Figure~\ref{fig:ads}(b) shows the change of the Mg-O bonding distance with the degree of coverage. Upon interaction with the surface, $d_{\rm Mg-O}$ increases from the initial 1.75 \AA\ in a free molecule to 2.01 \AA~for the full ML coverage. This value is lower than the value in bulk MgO (2.11~\AA),~\cite{Wyckoff1963} which is due to the pseudomorphicity with the Fe(001) surface. In the int-ot configuration, a minimum for 5/9 ML coverage is noted, which results from the previously mentioned lower symmetry of the MgO adsorbate structure. The 4\% lattice misfit between MgO and the Fe(001) surface causes the Mg atoms to be closer to the surface than O atoms by 0.11~\AA. Such a rumpling of the MgO layer closely reproduces the results of Jeon \textit{et al.}~\cite{Jeon2011}
Whereas the complete monolayer is pseudomorphic, for incomplete MgO coverage the interaction between Mg and O atoms on the surface causes the O atoms not to be located directly on-top of Fe atoms but to be slightly shifted toward the Mg atoms (Fig.~\ref{fig:fconf}).

The calculated height of the O atom above the Fe surface, $h_{\rm O}$, increases with the degree of coverage from the initial value of 1.90~\AA~for $\Theta=$1/9 to 2.24~\AA~for $\Theta=$1.  Our results also agree with the measured Fe-O distance, 2.20 $\pm$ 0.05~\AA, at the Fe(001)/MgO(001)  interface.~\cite{Wang2010} 
At 1~ML coverage, for the int-ot adsorption geometry, there is no incorporation of oxygen into the surface Fe layer. This configuration represents the sharp nonoxidized MgO/Fe interface, as discussed in the so-called Mg-rich condition.~\cite{Yu2006} 

\begin{table}[]
\caption{Relaxations $\Delta_{ij}$ (in \% of the bulk interplanar spacing) of the three top interlayer distances for the Fe(001) surface as a function of the MgO coverage in the int-ot configuration. Positive and negative values correspond to the expansion and contraction of the spacing, respectively.}
\label{tab:relax}
\begin{ruledtabular}
\begin{tabular}{cccc}
 Coverage (ML) & $\Delta_{12}$ & $\Delta_{23}$ & $\Delta_{34}$  \\ \hline
 0	   &	-2.1	  &	3.4	  &	0.7 \\
 1/9	   &	-1.2	  &	2.7	  &	0.4   \\
 2/9	   &	-0.8	  &	2.6	  &	0.4   \\
 3/9	   &	-0.3	  &	2.1	  &	0.4   \\
 5/9	   &	 0	  &	2.1	  &	0.6   \\
 7/9	   &	-0.6	  &	2.1	  &	0.6   \\
 9/9	   &	-2.1	  &	2.6	  &	0.9   \\
\end{tabular}
\end{ruledtabular}
\end{table}

Table~\ref{tab:relax} demonstrates that MgO adsorption in the int-ot configuration does not cause significant changes in relaxation of the interlayer distances at the Fe(001) surface. A weak derelaxation can be seen at fractional coverages. However, for 1 ML coverage, relaxations are nearly identical to the clean Fe(001) surface, indicating that the Fe(001) surface is stable against the adsorption of the MgO monolayer, when a sharp metal-oxide interface is formed. 

The interface structure is highly sensitive to the initial orientation of the MgO molecule with respect to the Fe surface. When the perpendicularly orientated MgO molecules approach the surface to form the O-int configuration, the O atom oriented toward the surface is incorporated into the Fe surface. For a complete MgO monolayer, the O atoms are 0.725~\AA~above the Fe(001) surface. In this case, formation of the Fe-O layer is possible. This would result in the oxidized MgO/Fe interface with an uncompensated surface layer of MgO. To make this case more realistic, we simulated O-rich conditions by providing additional oxygen atoms adsorbed in the on-top positions. This resulted in a stable surface MgO monolayer and a subsurface FeO layer. The FeO layer  showed only a small rumpling of 0.388~\AA, which corresponds to a nearly perfect oxidized MgO/FeO/Fe(001) interface, in agreement with other calculations.~\cite{Yu2006,Jeon2014} 

\begin{figure*}[]
\includegraphics[width=0.90\textwidth]{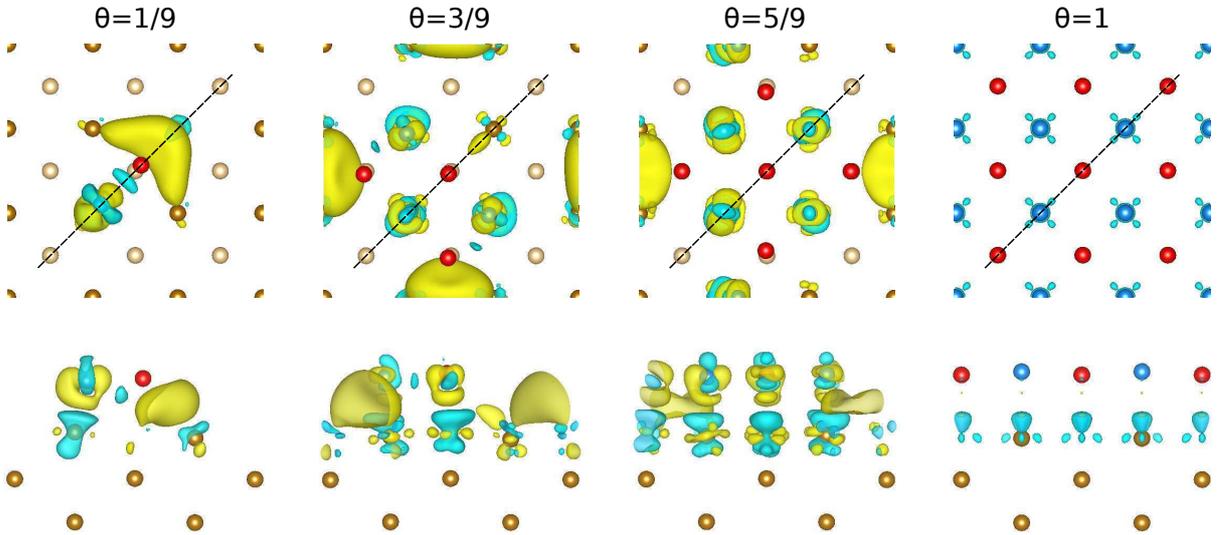}
\caption{(color online) Top and side (bottom panels) views of isosurfaces of the valence charge density difference, $\Delta\rho (\textbf{r})$, depending on the MgO coverage. Electron charge gain/loss is drawn in yellow/blue. Isosurface density level is 0.005 \textit{e}/bohr$^3$. Red, blue, and brown balls represent Mg, O and Fe atoms, respectively. The black lines in the top view panels mark the plane cuts for the side views shown in the lower row panels.}
\label{fig:chg}
\end{figure*}
 
\begin{figure}[]
\includegraphics*[width=8.0cm]{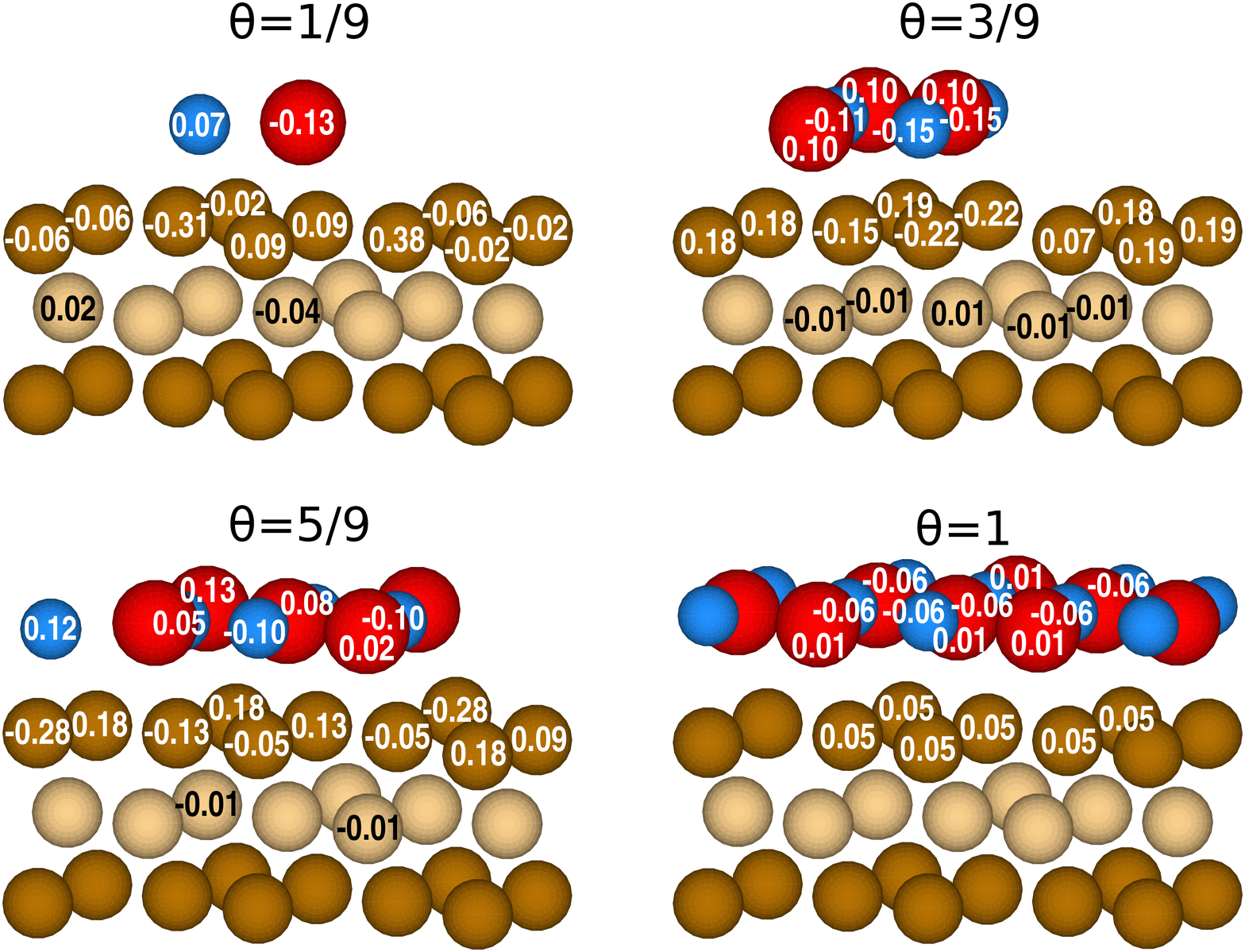}
\caption{(color online) Bader charge difference ($\Delta Q$) on the surface atoms of the top layers of the MgO/Fe(001) system induced by different degrees of MgO coverage.}
\label{fig:bader}
\end{figure}

The variations in the work function with changes in the coverage of MgO adsorbed in the int-ot configuration are plotted in Fig.~\ref{fig:ads}(c). The work function decreases gradually from 3.85 to 2.45 eV. This large reduction results from a difference between a small increase in the vacuum electrostatic potential and a pronounced shift of the Fermi energy of the MgO/Fe(001) system. 
The changes in work function with coverage reflect the changes in the electron charge-density distribution induced by MgO adsorption. Figure~\ref{fig:chg} diplays the isosurfaces of the electron density difference: 
\begin{displaymath}
 \Delta\rho (\textbf{r}) = \rho^{\rm MgO/Fe(001)}(\textbf{r}) - \rho^{\rm Fe(001)}(\textbf{r}) - \rho^{\rm MgO}(\textbf{r}) \,,
\end{displaymath}
where $\rho^{\rm MgO/Fe(001)}$ is the electron density of the Fe(001) surface with adsorbed MgO, $\rho^{\rm Fe(001)}$ is the electron density of the Fe(001) surface with the frozen atomic positions, and $\rho^{\rm MgO}$ is the electron density of the frozen MgO adsorbate. 
For a single MgO molecule ($\Theta=1/9$), the isosurfaces show the local electron charge accumulation in the space between the Mg and the neighboring Fe atoms. This means that MgO binding to the surface is dominated by the Mg-Fe bonding, which favors a horizontal alignment of the MgO molecule. 
In general, the plots for different degrees of coverage show that  the electron charge is drawn from the less negative Mg atoms (Pauling electronegativity =1.31) toward the more electronegative Fe atoms (electronegativity =1.83). This results in a negative contribution to the surface dipole barrier of the clean Fe(001) and the lowering of the work function. However, this effect is partially compensated by electrons drawn by the strongly electronegative oxygen ions (electronegativity =3.44). Consequently, the electrostatic dipole barrier changes relatively little (by $\approx0.4$ eV) in the range of coverage considered, and the work function lowering is mainly due to the shift of the Fermi level to higher energies with MgO coverage. 

More quantitative information about the charge transfer for each atom is provided by the analysis of the calculated Bader charge \cite{Henkelman2006} difference: 
\begin{displaymath}
 \Delta Q=Q^{\rm MgO/Fe(001)} - Q^{\rm Fe(001)}-Q^{\rm MgO} \, ,
\end{displaymath}
where $Q^{\rm MgO/Fe(001)}$ is the Bader charge of the atoms of the MgO/Fe(001) system,  $Q^{\rm Fe(001)}$ is the charge on the atoms of the frozen Fe(001) surface, and $Q^{\rm MgO}$ is the charge of the frozen MgO molecule. 
The largest charge difference occurs at the surface Fe layer atoms (Fig.~\ref{fig:bader}). The O atoms and Fe atoms located under the oxygen lose electron charge, which is transferred to the Mg atoms and Fe atoms away from the adsorbate. An exception occurs when the O atom has a lower number of neighboring atoms in plane (e.g., for $\Theta$=1/9 and $\Theta$=5/9). In this case, oxygen gains electron charge. Much less electron transfer can be observed when the MgO monolayer is completed. The analysis of the total Bader charge shows transfer from oxygen to the Fe surface, which increases with the coverage, except for 4/9 and 5/9 ML. It can reduce the surface dipole barrier and consequently decrease the work function.

\subsubsection{Electronic and magnetic structure}

\begin{figure}[]
\includegraphics*[width=0.46\textwidth]{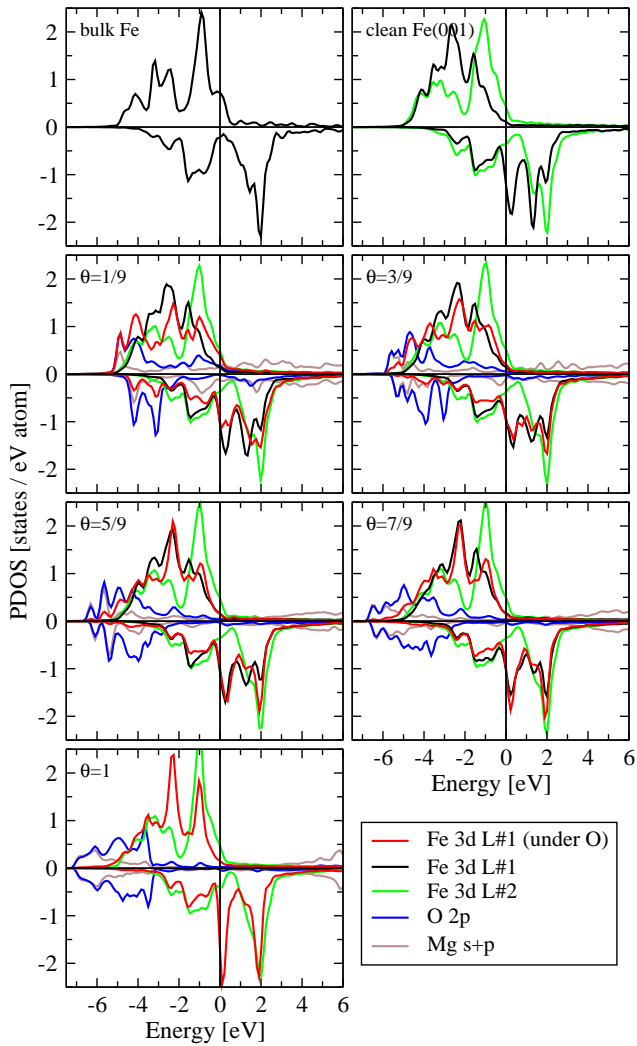}
\caption{(color online) Partial densities of states (PDOSs) of the clean Fe(001) surface and for different degrees of MgO coverage adsorbed in the most stable int-ot configuration. Positive and negative densities correspond to the spin-up and spin-down states, respectively. Energy zero corresponds to the Fermi level. The plots for Mg states are multiplied by 5.}
\label{fig:pdos}
\end{figure}

The partial, layer-resolved densities of states (PDOSs) for the considered system are presented in Fig.~\ref{fig:pdos}. For the clean Fe(001) surface, the changes in the PDOSs with respect to the bulk are practically limited to the atoms of the topmost Fe layer.   
For the first surface Fe layer, the PDOSs were plotted separately for the Fe atoms away from the O atoms and for those binding with the O atoms. In the entire range of coverage, the former are nearly unaltered compared with the Fe $3d$ states of the topmost layer of the clean Fe surface. The main changes are observed in the PDOSs of the Fe atoms located under the oxygen. 
This effect is particularly visible for low coverages. The proximity of the O atoms contributes to the additional peaks in the Fe $3d$ states due to hybridization with the O $2p$ states. The additional peaks disappear with the increasing coverage. For the full MgO monolayer, the $3d$ PDOS of the topmost Fe atoms is similar to that for the clean surface atoms, but the PDOS peaks are sharper and higher. With increasing coverage, the shape of the O $2p$ PDOS changes and the states are shifted to lower energies, which reflects the stronger MgO binding to the iron substrate.

\begin{figure}[]
\includegraphics*[width=6.0cm]{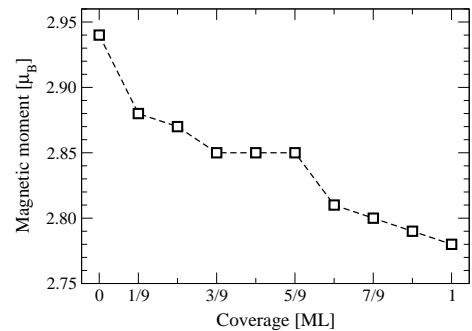}
\caption{Coverage dependence of the average magnetic moment on the Fe atoms in the topmost surface layer.}
\label{fig:moments}
\end{figure}

The presence of the Mg and O adatoms on the surface induces changes in the magnetic moment of the Fe atoms of the topmost surface layer.
The values of the local magnetic moment vary between 3.00$\mu_{B}$ for Fe atoms without contact with an adsorbate and 2.59$\mu_{B}$ for Fe atoms adjacent to the MgO molecules. The average magnetic moment of the surface Fe atoms plotted in Fig.~\ref{fig:moments} as a function of MgO coverage gradually decreases with the coverage, from 2.94$\mu_{B}$ for the clean surface to 2.78$\mu_{B}$ for 1 ML of MgO.  The latter value is 5\% larger than the value previously reported for 1 ML of MgO on Fe(001).\cite{Yu2006}
Generally, the proximity of oxygen enhances the magnetic moment of iron,~\cite{Blonski2005,Jeon2014} which means that nonmagnetic Mg atoms are responsible for weakening the magnetism on the Fe surface. As a result of the Mg-atom proximity, the Fe $3d$ majority states are shifted toward higher energy. Their occupancy decreases, resulting in the reduction of the magnetic moment of Fe. The adsorption of MgO has a negligible effect on the magnetic moments of the second Fe layer, which are nearly unaltered compared to the corresponding moments of the clean Fe(001) surface. Hybridization of the Fe $3d$ states and O $2p$ states induces a small magnetic moment on the O atoms, which changes with coverage. The magnetic moment ranges between 0.17$\mu_{B}$ and 0.03$\mu_{B}$ for $\Theta=1/9$ and 1 ML of coverage, respectively. A small magnetic moment of 0.01--0.03$\mu_{B}$ also appears at the Mg atoms.

\section{Summary and Conclusion}

We have presented the results of a DFT study of the adsorption of MgO molecules on a Fe(001) surface for coverage varying from 1/9 to 1 ML. The most stable configuration of the adsorbed MgO molecule is with the Mg atoms in the interstitial site and the O atoms on-top of Fe site (int-ot configuration) for the molecule aligned parallel to the surface. The configuration with Mg at the interstitial and O at the bridge site (int-bri) that is preferred at the initial adsorption stages becomes unstable for higher degrees of coverage. This confirms the \textit{a priori} assumption that the int-ot configuration of the adsorbed MgO on the Fe(001) surface~\cite{Dugerjav2011,Meyerheim2001} is the preferred arrangement. Under circumstances that simulate the adsorption of stoichiometric MgO, we show a preference for the sharp MgO/Fe interface formation without oxidation of the topmost Fe layer.

Our results also show that oxidation of the Fe(001) surface cannot be excluded during the adsorption of MgO molecules aligned perpendicular to the Fe surface with the O atoms facing the substrate. The analysis of the adsorption energy shows that this configuration is less probable in stoichiometric conditions than the adsorption of MgO molecules aligned parallel to the surface. Nevertheless, in this particular O-int configuration, the O atoms significantly approach the surface and an FeO-like layer is formed. This is consistent with many experimental works~\cite{Oh2003,Tusche2006,Mlynczak2013}, reporting the appearance of the FeO phase at the interface in MgO/Fe systems.

\begin{acknowledgments} 
This work was supported by the National Science Center, Poland (Grant No.\ UMO-2011/02/A/ST3/00150). Computer time from the Interdisciplinary Centre for Mathematical and Computational Modelling (ICM) of the Warsaw University (Project No.\ G44-23) is gratefully acknowledged. The research was performed in the framework of the Marian Smoluchowski Krakow Research Consortium - Leading National Research Centre (KNOW), which is supported by the Polish Ministry of Science and Higher Education.
\end{acknowledgments}

\bibliography{First_principles_study_of_the_adsorption_of_MgO_molecules}{}

\end{document}